\documentclass[twocolumn,showpacs,preprintnumbers,amsmath,amssymb]{revtex4}
\usepackage{graphicx}
\usepackage{amsmath}
\usepackage{color}

\begin{document}
\title{Triggering rogue waves in opposing currents}

\author{Miguel Onorato}
\affiliation{Dipartimento di Fisica Generale, Universit{\`a} degli Studi di Torino, Via Pietro Giuria 1, 10125 Torino, Italy}
\affiliation{INFN, Sezione di Torino, Via Pietro Giuria 1, 10125 Torino, Italy}

\author{Davide Proment}
\affiliation{Dipartimento di Fisica Generale, Universit{\`a} degli Studi di Torino, Via Pietro Giuria 1, 10125 Torino, Italy}
\affiliation{INFN, Sezione di Torino, Via Pietro Giuria 1, 10125 Torino, Italy}

\author{Alessandro Toffoli}
\affiliation{Faculty of Engineering and Industrial Sciences, Swinburne 
University of Technology, P.O. Box 218, Hawthorn, 3122 Vic., Australia}

\date{\today}

\begin{abstract}
We show that rogue waves can be triggered naturally when a stable wave train enters a region of an opposing current flow. 
We demonstrate that the maximum amplitude of the rogue wave depends on the ratio between the current velocity, $U_0$, and the wave group velocity, $c_g$. We also reveal that an opposing current can force the development of rogue waves in  random wave fields, resulting in a substantial change of the statistical properties of the surface elevation. 
The present results can be  directly adopted in any field of physics in which the focusing Nonlinear Schrodinger equation with 
non constant coefficient is applicable. In particular, nonlinear optics laboratory experiments are natural candidates for verifying experimentally our results.
\end{abstract}
\maketitle
In the ocean, rogue waves are often observed in regions characterized by strong currents like 
the Gulf Stream, Agulhas Current and the Kuroshio Current  \cite{lavrenov1998wave,malloryabnormal}. 
Several ship accidents have 
been reported in these regions as being due to the impact with very large waves. One of these occurred in February 1986 to the SS Spray, which was travelling along the East coast of the USA. The ship was hit by a wave with a height of approximately 17 $m$ (estimated by eyes from the deck of the ship), which was the second of a system of three consecutive large waves, commonly  known as the three sisters.
This particular wave system is usually observed in the nonlinear stages of the modulational instability process.
Such instability was discovered in the late sixties independently by Zakharov \cite{zakharov68} and Benjamin and Feir \cite{BF67} 
(an interesting and stimulating review on the subject can be found in \cite{west1981}).
The theory is based on the linear stability analysis of a plane wave and predicts that a
small perturbation may grow exponentially when $\varepsilon N>1/\sqrt2$,
where $\varepsilon=k_0 A_0$ is the steepness of the plane wave, with $k_0$ its wave number and $A_0$ its amplitude; $N=\omega_0/\Delta \Omega$ is the number of waves under the modulation, with $\omega_0$ the angular frequency corresponding to the wave number $k_0$ and $\Delta \Omega$ the angular frequency of the modulation. 

The nonlinear stages of 
the modulational instability are described by the exact breather solutions of the Nonlinear Schr{\"o}dinger (NLS) equation \cite{akhmediev1987exact,kuznetsov1977solitons,ma1979perturbed}.
Breathers are coherent structures that oscillate in space or time and
have the peculiarity of changing their amplitudes as they propagate. 
They
can grow up to a maximum of 3 times their initial amplitude and have been considered in various fields of physics as a plausible object that describes the formation of rogue waves
\cite{dysthe99,osborne00}. Such solutions 
have also been observed in fully nonlinear simulations of 
the Euler equation, \cite{dyachenko2008formation}.
Breather solutions may also exist embedded in random waves \cite{onorato01} and can affect the probability density function of the surface elevation and wave height distribution \cite{onoratoetal04,morietal07}. 
 We mention here that the concept of rogue waves is 
rapidly expanding to other disciplines such as nonlinear optics and condensed matter
(see, for example, 
\cite{solli2007optical,kibler2010peregrine,kharif2009rogue,
bludov2009matter,ruban2010rogue,montina2009non}).

In general ocean waves are characterized by a small value of $\varepsilon N$.
Wind seas, which are waves forced by the local wind field, have moderately large steepness but the spectral band-width only rarely allows for a large number of 
the product  $\varepsilon N$ that satisfies the instability criterion. On the other hand, swells, i.e. long crested waves that have  moved out of the generating region, are characterized by a narrow spectrum (i.e. large $N$)
both in angle and frequency, but they are not particularly steep. Hence, breathers (rogue waves)
are fortunately rare objects in the ocean. However, if swells enter into a current, their properties
can change and, as we will show in this Letter, breathers solutions can be 
naturally triggered.

Our analysis is based on the computation of a modified Nonlinear Schroedinger 
equation, recently derived in  \cite{hjelmervik2009freak}, that accounts for a  current, $U=U(x)$, assumed to be small with respect to the wave phase velocity $c_p$: $U/c_p=O(\varepsilon)$.
 The derivation requires also that the current is a slowly varying function of the spatial coordinate, i.e., $1/(\Lambda k_0)=O(\varepsilon)$, with $\Lambda$ the entry length of the current (typical space scale over which the current changes).
The equation describes the evolution in space of the wave envelope $A=A(x,t)$. 
In dimensional variables it takes the following form:
\begin{equation}
\begin{split}
&
\bigg[\frac{\partial A}{\partial x}+
\frac{1}{ c_g}
\left(1-
\frac{3}{2}\frac{U}{ c_g}\right)\frac{\partial A}{\partial t}\bigg]+
i \frac{k_0}{\sigma_0^2}\frac{\partial^2 A}{\partial t^2}
+ 
i k_0^3 |A|^2A=  
  \\
 &=-\frac{1}{2  c_g}\frac{dU}{d x}A-
i k_0\frac{U}{ c_g} \left(
1-\frac{5}{4}\frac{U}{ c_g}\right) A,
\end{split} \label{eq:trulsen}
\end{equation}
with $c_g$ the group velocity, 
 $k_0$ the wave number of the carrier wave and $\sigma_0=\sqrt{g k_0}$, with $g$ the gravity acceleration.
We have found that equation  (\ref{eq:trulsen}) can be reduced to the standard NLS equation with variable coefficients by applying the following transformations:
\begin{equation}
A=B\exp\left[- \int_0^xi k_0\frac{U}{ c_g} \left(
1-\frac{5}{4}\frac{U}{ c_g}\right)-\frac{1}{2 c_g} \frac{dU}{dX}dX\right],
\end{equation}
\begin{equation}
x'=x,\;\;\;\;\;\;\;
t'=t-\int_0^x \frac{1}{V(X)}dX,
\end{equation}
after which,  equation (\ref{eq:trulsen})  assumes the form
\begin{equation}
\begin{split}
&
\frac{\partial B}{\partial x}+i \frac{k_0}{\sigma_0^2}\frac{\partial^2 B}{\partial t^2}+ 
i \beta(x)
|B|^2B=0 \label{eq:NLS_fin}
\end{split}
\end{equation}
where
\begin{equation}
\begin{split}
&\frac{1}{V(x)}=\frac{1}{ c_g}\bigg(1
-\frac{3}{2}\frac{U}{ c_g}\bigg), \;\;
\beta(x)=
k_0^3 exp\left[-\frac{\Delta U}{ c_g} \right],
\end{split} \label{eq:coeff}
\end{equation}
with $\Delta U(x)=U(x)-U(0)$;
primes have been omitted for brevity.
Equation (\ref{eq:trulsen}) does not preserve the energy, 
which can be shown to change in space as follows:
$E(x)=\int |A|^2dt= \exp\left[-\Delta U/ c_g \right]\int |B|^2dt$.
As the waves enter into a current, their wave height increases 
if $\Delta U<0$ (opposing current) and decrease if $\Delta U>0$
(co-propagating current).

In the absence of a current, U=0, equation (\ref{eq:trulsen}) admit
breather solutions, \cite{akhmediev1987exact},
whose maximum amplitude 
reached during the evolution of the wave group is:
\begin{equation}
{\frac {A_{max}} {A_0} = 1 + 
2 \sqrt{ 1-\bigg(\frac{1} {\sqrt{2}\varepsilon N}\bigg)^2 }}.
\label{amax3_mig}
\end{equation}
For $ \varepsilon N < 1/\sqrt{2}$, wave groups are stable; for $\varepsilon N > 1/\sqrt{2}$, the amplification factor reaches the maximum value of three, which corresponds to the Peregrine soliton,
observed experimentally recently in nonlinear optical fibers \cite{kibler2010peregrine}.

\begin{figure}[h]
\includegraphics[width=8cm]{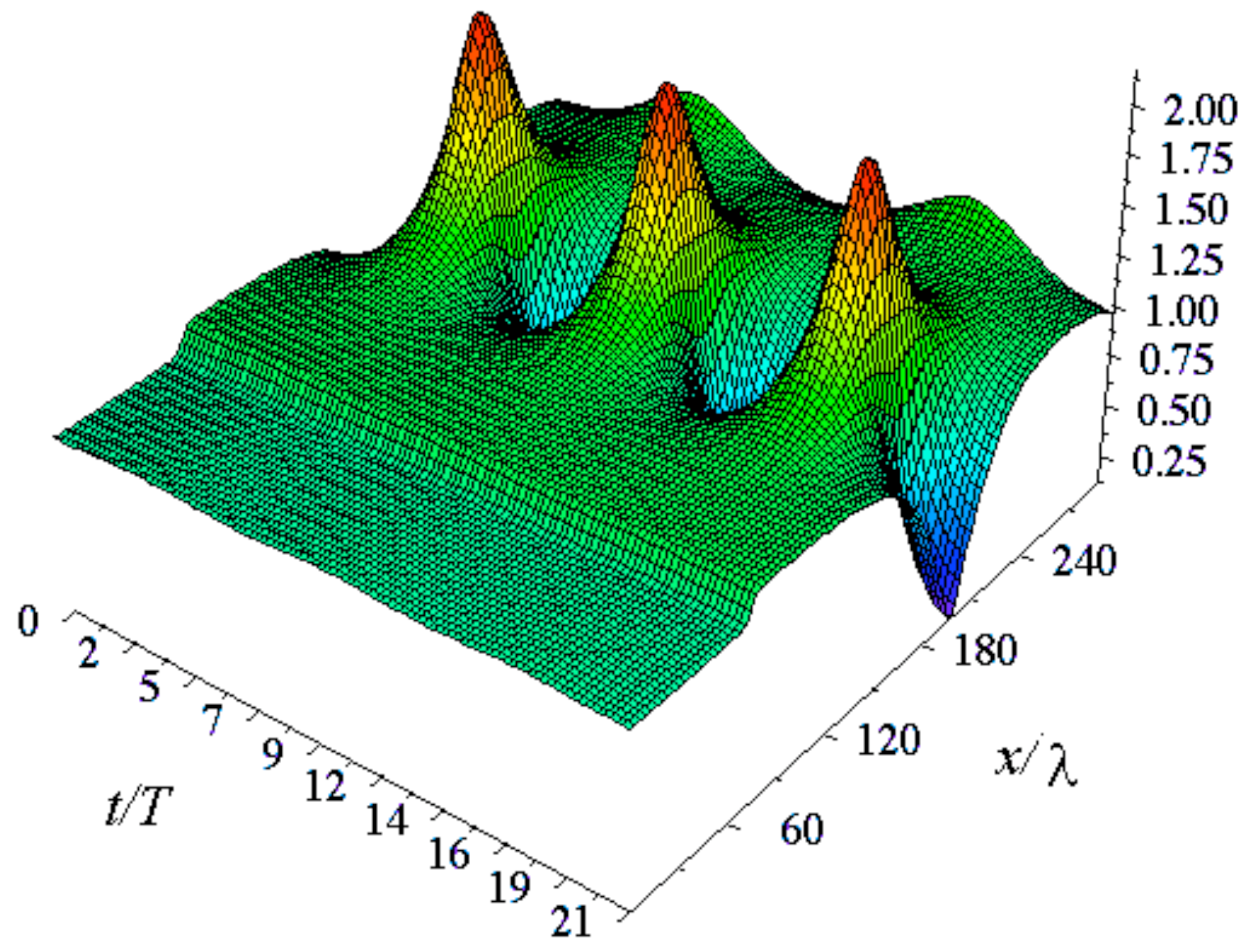}
\caption{Evolution in space and in time, normalized by the wave length and period respectively,  of the wave envelope. The wave propagates for 60 wavelengths before entering into the opposing current of velocity $U_0=-0.25 c_g$. The effect of the current is not only
to increase the energy (amplitude) of the wave but also to trigger the formation of a rogue wave.}
\label{fig:breather}
\end{figure}

\begin{figure}
\includegraphics[width=7cm]{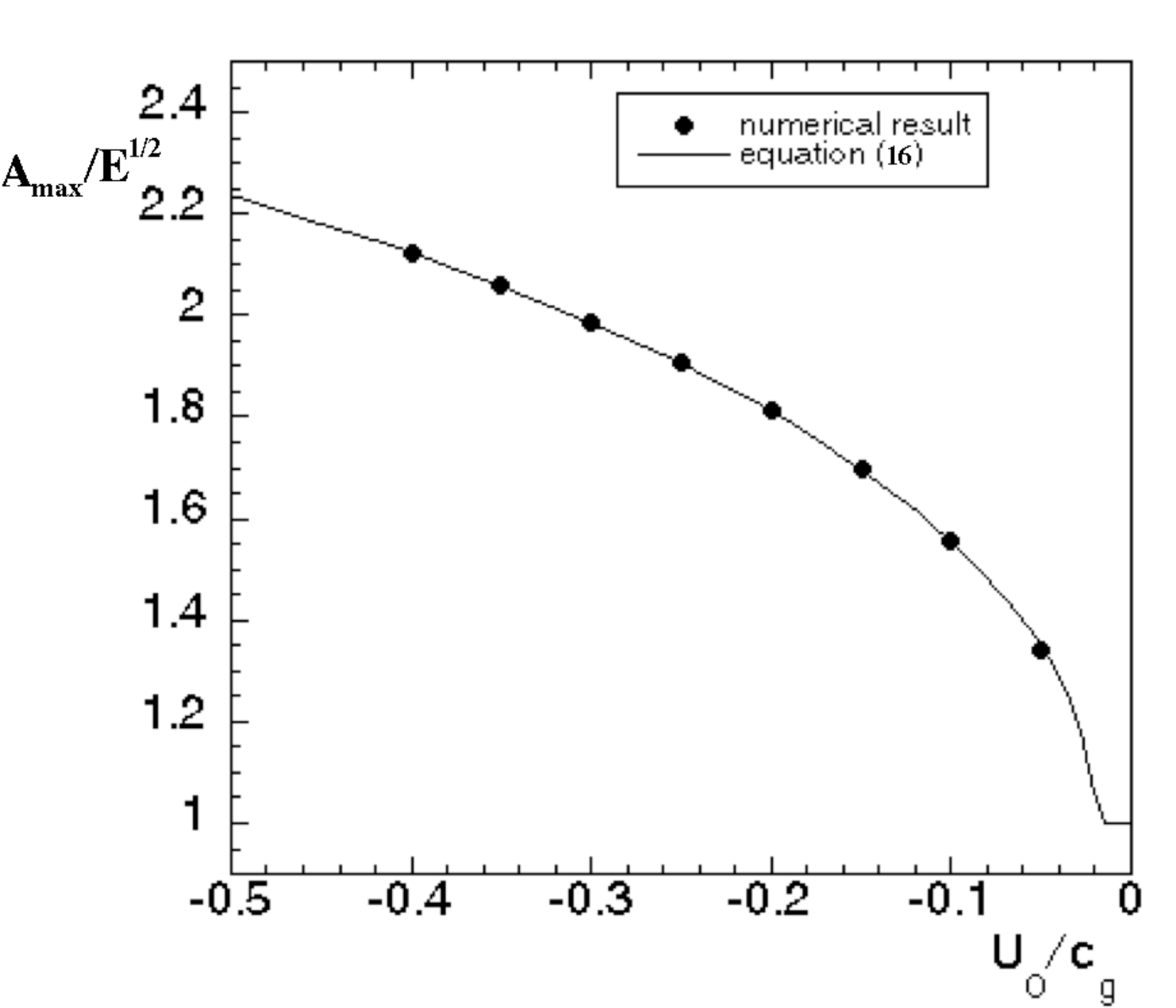}
\caption{The normalized maximum amplitude as a function of  $U_0/c_g$.
The dots are the results from the numerical simulations and 
the solid line corresponds to equation (\ref{eq:amax3_mig_n}).}
\label{fig:breather_max}
\end{figure}

We will consider the evolution of a wave train initially in a region of zero current  propagating into a stationary current characterized by  
an entry length $\Lambda$. We have used in our computation the following simple mathematical expression for the current:
\begin{equation}
  U(x) = \left\{
  \begin{array}{l l }
    0 & \quad \text{if                       } x < x_0 \\ 
    U_0 \sin^2\left[\frac{\pi}{2\Lambda}{(x-x_0)}\right] & \quad \text{if   } x_0\le x< x_0+\Lambda \\
   U_0 & \quad \text{if                       } x \ge x_0+\Lambda \\
  \end{array} \right.
  \label{eq:current}
\end{equation}
Here $x$ is the coordinate along which the waves propagate and $U_0$ is the asymptotic value (large $x$) of the current  that can be either positive or negative for co-propagating or opposing currents,
respectively. For such a current field,  we show a first example of the formation of a rogue wave  
 in an opposing current in  Figure \ref{fig:breather}.
 As initial conditions for our numerical experiments,  we have considered a perturbed plane wave 
 with $\varepsilon=0.1$ and $N=7$, which are typical values for an energetic swell; in terms of dimensional quantities, we can imagine a wave system characterized by a period of 10 seconds (0.1 $Hz$). In absence of current, the wave train is stable and no modulational instability is observed. 
 After 60 wavelengths of propagation, the wave group enters into a current characterized by $U_0/c_g=-0.2$
 (opposing current) and $\Lambda=10 \lambda$. At $x/\lambda=60$ the whole envelope grows in amplitude because the energy $E(x)=\int |A(x,t)|^2dt$  changes, in the presence of a current gradient. Subsequently the envelope starts being strongly modulated,
 reaching approximately two times its local standard deviation, $\sqrt{E(x)}$. The plot shows a 
 clear example of formation of a rogue wave starting from a stable plane wave.
 We have performed a systematic study on the dependence of the maximum amplitude (divided 
by $\sqrt{E(x)}$)
reached by the envelope
as a function of the ratio $U_0/c_g$ which was varied from 0.1 to 0.4
(stronger current may result in wave blocking phenomena and 
wave breaking).
The results are shown  in Figure \ref{fig:breather_max} where the dots corresponds 
to our numerical results.
%
%
The plot indicates that the normalized
maximum amplitude increases with increasing $|U_0|$, evidencing that the originally stable plane 
wave has been transformed into a breather in the presence of a current.  
In the figure we also include as a solid line the heuristic prediction based on the following 
equation:
\begin{equation}
\frac {A_{max}} {\sqrt{E(x_0+\Lambda)}} = 
1 + 2 \sqrt{ 1-\bigg(\frac{\exp[U_0/(2 c_g)]} {\sqrt{2}\varepsilon N}\bigg)^2 }.
\label{eq:amax3_mig_n}
\end{equation}
Equation (\ref{eq:amax3_mig_n}) is a modification of the 
exact relation (\ref{amax3_mig}). The rationale behind this 
prediction is that the coefficient in front of the nonlinear term 
in equation (\ref{eq:NLS_fin}), once written in non-dimensional form, is the same as the 
one without current except for the exponential factor (see the coefficient $\beta(x)$ in (\ref{eq:coeff})). 
As shown in figure \ref{fig:breather_max},  
the numerical results are in excellent agreement with the prediction. 

\begin{figure}
\begin{center}
\includegraphics[width=7cm]{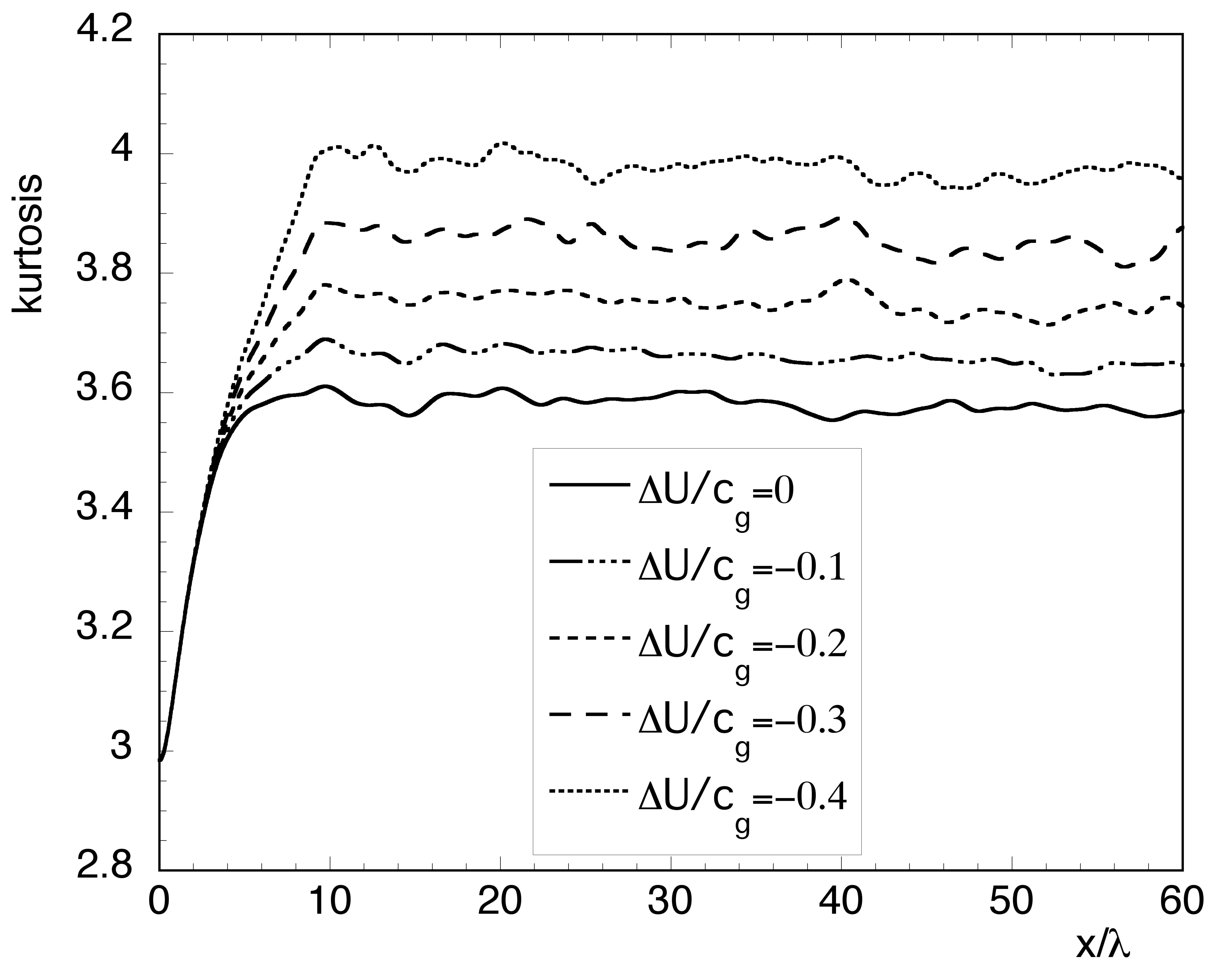}
\caption{Evolution of the kurtosis as a function of the normalized distance $x$ for  different simulations corresponding to 
different values of $\Delta U/c_g$. For larger values of the $\Delta U/c_g$ there is a clear indication of stronger deviations from Gaussian statistics.  }
\label{fig:kurt}
\end{center}
\end{figure}

Once established the possibility that an opposing current may 
trigger unstable modes, it is of interest to understand whether the statistical properties of the surface elevation and in particular the occurrence of extreme events change as random wave trains 
propagate into the current. We concentrate our analysis on the 
kurtosis, $\kappa$, i.e. the fourth order moment of the probability density function of the surface elevation
estimated as $\kappa={\langle \eta^4\rangle}/{\langle \eta^2\rangle^2}$,
where $\langle...\rangle$ stands for the ensemble average. 
In the presence of a current, an analytical estimation of the kurtosis is a rather difficult task because of the nonlinearity of the problem. Therefore, we perform  direct  numerical simulations of the modified NLS equation  with initial conditions characterized by the following bell-shape spectrum for the envelope $A$:
\begin{equation}
P(\omega)=\frac{E}{ \Delta\Omega \sqrt{2 \pi}}
\exp{\left[-\frac{\omega^2}{2\Delta\Omega^2}\right]}
\end{equation}
with $\Delta \Omega$  the standard deviation (the width of the spectrum).
The phases are considered randomly distributed in the 
$[0,2\pi)$. Numerical simulations are computed on a grid of 1024 points and 600 realizations 
have been performed. The wave steepness $\sqrt 2 k_0 \sqrt { E}$ was selected equal to 0.15 and $\Delta\Omega/\sigma_0=0.2$.
The kurtosis is therefore estimated first as a time average and then the resulting value is averaged over the 
ensemble. The current is characterized by $x_0=0$ and 
$\Lambda=10 \lambda$ and different values of $U_0/c_g$ are considered.
In Figure \ref{fig:kurt}, the kurtosis is presented 
as a function of $x/\lambda$.
\begin{figure}
\begin{center}
\includegraphics[width=7cm]{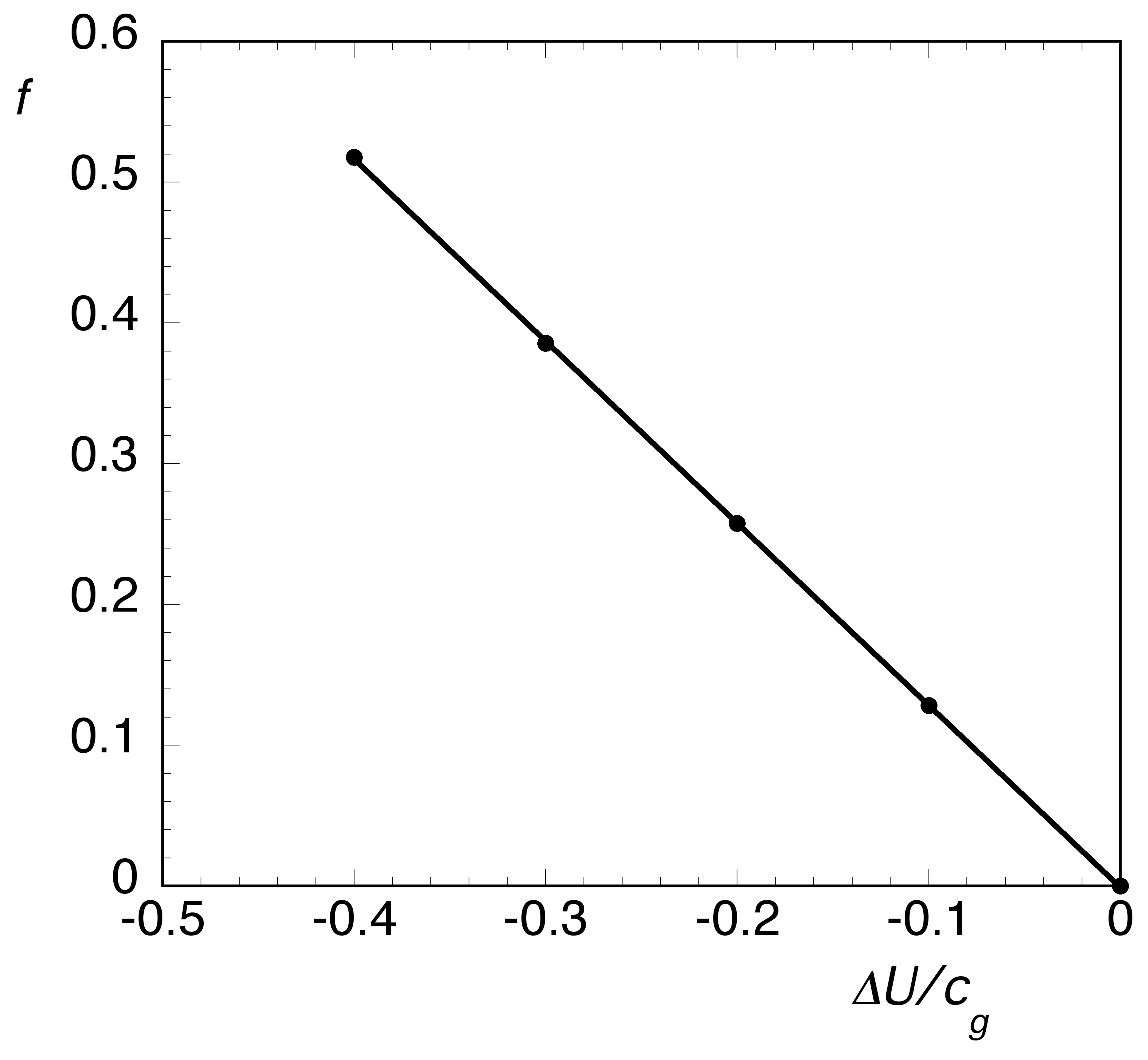}
\caption{ $f(\Delta U/c_g)=\ln[(\kappa-3)/(a \pi  BFI^2/\sqrt3)]$ as a function of $\Delta U/c_g$. Dots 
are the results from the numerical computation and the solid line is obtained by a linear fit.}
\label{fig:fit}
\end{center}
\end{figure}
The figure clearly indicates a dependence of the kurtosis on the 
ratio between the current increment and the wave group velocity. This result is to some extent consistent with recent laboratory experiments, which investigated the evolution of mechanically generated, random wave fields over a partially opposing current \cite{toffoli11}.
In the absence of a current, as shown in \cite{janssen03}, the kurtosis 
depends on the square of the Benjamin-Feir Index (BFI):
\begin{equation}
BFI=\frac{\sqrt{E}k_0}{\Delta \Omega/\sigma_0},
\end{equation}
which is the ratio between the nonlinear and the linear coefficients in the NLS equation properly written in nondimensional form see \cite{onorato01}.
 In the presence of the current, 
 we assume the following dependence on the current velocity:
\begin{equation}
\kappa=3+a \frac{\pi}{\sqrt3}  BFI^2
\exp\left[-b\frac{U_0}{c_g}\right] , 
\label{eq:fitting}
\end{equation}
%
where $a$ and $b$ can be  determined {\it a posteriori}
 from the numerical simulations. For  $a=1$ and $b=1$,  the 
kurtosis estimated through (\ref{eq:fitting}) corresponds to an analytical result that can be obtained by a quasi-gaussian approximation under the hypothesis that the wave spectrum and the current field are slowly varying in space. The derivation follows the one presented in \cite{janssen03} performed in the absence of current. We mention here that our aim is not to 
establish quantitatively  the validity of the closure model but to understand how the opposing current influence
the statistics of the waves.
\begin{figure}[h!]
\begin{center}
\includegraphics[width= 7cm]{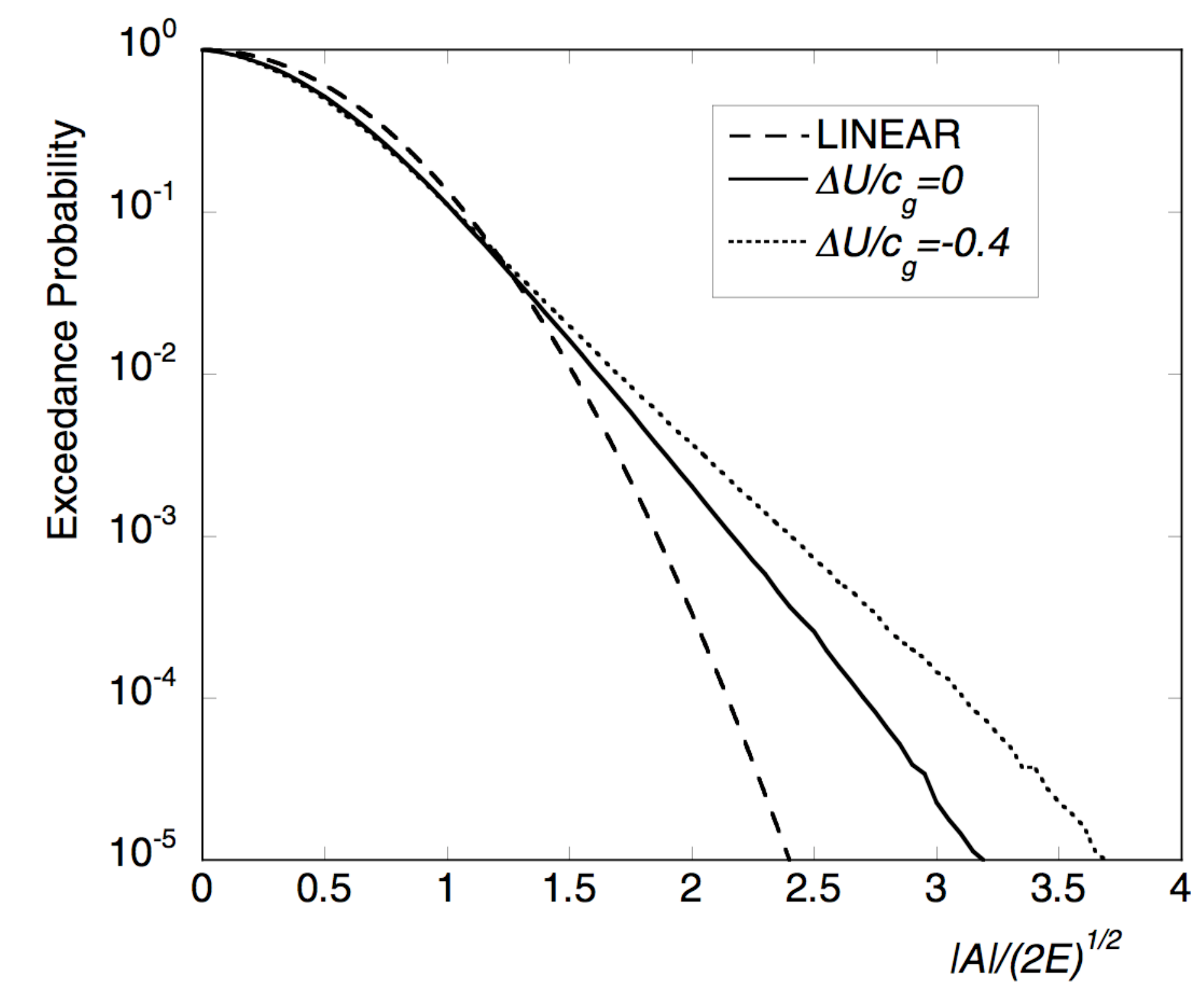}
\caption{The probability of exceedance for $|A|/ \sqrt {2E}$ for the case of a linear system (dashed line), nonlinear system without current (solid line) and nonlinear system in the presence of an opposing current (dotted line). }
\label{fig:pdf}
\end{center}
\end{figure} 
In order to verify the exponential factor in (\ref{eq:fitting}), we first 
compute from the initial conditions the $BFI$, then the asymptotic (large $x$) value of the kurtosis is measured 
as an average value of the kurtosis over the last 20 wavelengths (see 
Figure \ref{fig:kurt}); at last we determine the coefficient $a$ 
from the simulation with $\Delta U=0$ (the resulting value is $a=1.133$).  In Figure \ref{fig:fit} we show the quantity $f(\Delta U/c_g)=\ln[(\kappa-3)/(a BFI^2 \pi/\sqrt3)]$ as a function of $\Delta U/c_g$. 
The points from the simulation lie on a straight line which indicates 
that the assumed exponential dependence is consistent with our numerical simulations.
In the figure we also show the linear fit, where the slope is $b=1.29$.
Moreover, we consider the exceedance probability 
 (defined as 
$\int_x^{\infty}p(x') dx'$, with $p(x)$ the probability density function)
for the envelope computed at $x$=60 $\lambda$ for the case 
of $\Delta U/c_g=0$ and  $\Delta U/c_g=-0.4$. 
The probability of occurrence of rogue waves increases notably as
the  opposing current is stronger. 
Results are presented in Figure \ref{fig:pdf} where we have also plotted the 
exceedance probability for Rayleigh distribution which is the one 
estimated for a linear process. 
In \cite{hammani2010emergence} the appearance of rogue waves in an NLS equation with third order dispersion has been discussed in terms of the ratio between the nonlinear and linear part of the Hamiltonian.  In our case such ratio, proportional to the $BFI$, calculated for the initial condition is equal to 0.3; our simulations are in the intermittent-like rogue wave regime, they appear and disappear erratically.

Ocean swells are in general not very steep and wave packets are stable  in terms of modulational instability. However, we have shown that breathers  may be triggered when 
swells enter into a region of opposing current. This is an important result that should be kept
in mind when ships navigate in the Gulf Stream or in the Agulhas Current or Kuroshio
Current in the presence of a opposing waves. Indeed, such currents may  reach velocities up to 1.5 meter per second and for a 
group velocity corresponding to waves of period equal to 10 second (a typical condition during storms), the ratio
$\Delta U/c_g$ is of the order of 0.2, large enough to trigger a 
dangerous rogue wave. We underline that this is 
completely different  process from the development of a caustic,  a pure linear mechanism
 \cite{WHITE,lavrenov98,Heller2008}. 
  From a physical point of view, the mechanism of formation of rogue waves can be summarized as follows: an initial wave whose perturbation is stable in terms of the modulational instability may become unstable in the presence of a current because of the a shift of the modulational instability band. The modulational instability of the wave thus leads to a triggering of the rogue wave. 
The results presented in this Letter,  even though presented in the oceanographic context, may also apply to the nonlinear optics physics where the nonlinearity is provided by the type of the
material. As shown in equation (\ref{eq:NLS_fin}), the effect of the current is to change the coefficient 
of the nonlinear term in the NLE equation. 
 New experiments characterized by materials that change
their nonlinear properties in space could be easily performed and 
the predictions of the present work could be verified.
{\bf Acknowledgments} K. Trulsen and Al Osborne are acknowledged for discussions. This work has been funded by EU, project EXTREME SEAS (SCP8-GA-2009-234175). A.T. was supported by the Australian Research Council and Woodside Energy Ltd Linkage project LP088388. The present work has started when M.O. was visiting the Swinburne University of Technology under the program Visiting Professor Award Scheme of Swinburne University of Technology.
\bibliography{references}
\end{document}